\documentclass[aps,prd,nopacs,twocolumn,nofootinbib,notitlepage,longbibliography,superscriptaddress]{revtex4-1}

\usepackage{placeins}
\usepackage{tabularx}
\usepackage{array}
\usepackage{subfig}
\usepackage{amsmath,amssymb,amsfonts,bm}
\usepackage{graphicx}
\usepackage[usenames,dvipsnames]{xcolor}
\usepackage[normalem]{ulem}
\usepackage[colorlinks=true,hyperfootnotes=true, linkcolor=BrickRed, citecolor=RoyalBlue, urlcolor=magenta]{hyperref}
\usepackage{cases}

\def\dd{\mathrm{d}}

\newcommand{\bc}{\begin{center}} 
\newcommand{\ec}{\end{center}}
\newcommand{\be}{\begin{equation}} 
\newcommand{\ee}{\end{equation}}
\newcommand{\ba}{\begin{eqnarray}}
\newcommand{\ea}{\end{eqnarray}}

\begin{document}

\title{Non-canonical 3-form dark energy}

\author{Vitor da Fonseca}
\email{vdafonseca@alunos.fc.ul.pt}
\affiliation{Instituto de Astrof\'isica e Ci\^encias do Espa\c{c}o,\\ 
Faculdade de Ci\^encias da Universidade de Lisboa,  \\ Campo Grande, PT1749-016 
Lisboa, Portugal}
\author{Bruno J. Barros}
\email{bjbarros@fc.ul.pt}
\affiliation{Instituto de Astrof\'isica e Ci\^encias do Espa\c{c}o,\\ 
Faculdade de Ci\^encias da Universidade de Lisboa,  \\ Campo Grande, PT1749-016 
Lisboa, Portugal}
\author{Tiago Barreiro}
\affiliation{Instituto de Astrof\'isica e Ci\^encias do Espa\c{c}o,\\ 
Faculdade de Ci\^encias da Universidade de Lisboa,  \\ Campo Grande, PT1749-016 
Lisboa, Portugal}
\affiliation{ECEO, Universidade Lus\'ofona de Humanidades e 
Tecnologias, Campo Grande, 376,  PT1749-024 Lisboa, Portugal}
\author{Nelson J. Nunes}
\affiliation{Instituto de Astrof\'isica e Ci\^encias do Espa\c{c}o,\\ 
Faculdade de Ci\^encias da Universidade de Lisboa,  \\ Campo Grande, PT1749-016 
Lisboa, Portugal}

\date{\today}

\begin{abstract}
In this study, we meticulously construct a 3-form Lagrangian designed to mimic the dynamics of both dust matter in the past and  dark energy driving the acceleration in the present era. A dynamical systems approach is used to investigate the underlying behavior of the cosmological background. By investigating the influence of the potential slope and initial conditions on the dynamical solutions, we identify distinct viable scenarios capable of replicating a De Sitter universe in the present epoch. An intriguing aspect of the model is the existence of solutions describing multiple inflationary phases in which the 3-form self-interacting potential decays rapidly.
\end{abstract}

\maketitle

\section{Driving the late-time acceleration with a 3-form}\label{Sec::intro}

In the literature, the predominant method to endow the cosmological constant with dynamical properties is undoubtedly with the use of scalar fields \cite{Wetterich:1994bg, Zlatev:1998tr,Chiba:1999ka,dePutter:2007ny}. In the most common scenario, the scalar field typically undergoes dynamical evolution during early epochs, with its potential eventually ensuring dominance in later periods, consequently leading to a late-time accelerated solution \cite{SupernovaCosmologyProject:1998vns, SupernovaSearchTeam:1998fmf, Caldwell:1997ii,Copeland:2006wr}. An appealing property of scalar field dark energy models is the existence of scaling solutions that have the potential to deal with shortcomings of the cosmological constant \cite{Copeland:1997et,Steinhardt:1999nw,Barreiro:1999zs,Peebles:2002gy}. In particular, they are expected to address the extreme fine-tuning of the initial conditions and to explain the transition to dark energy domination, although the parameters of the scalar field still require some fine-tuning. While these theories are diversified and continue to hold a prominent position in theoretical models of dark energy \cite{Armendariz-Picon:2000nqq,Armendariz-Picon:2000ulo,Carroll:2003st,Singh:2003vx,Sami:2003xv,Amendola:1999er,Khoury:2003aq,Khoury:2003rn,Kamenshchik:2001cp,Bento:2002ps}, alternative possibilities have emerged, each with its own compelling features \cite{Brax:2017idh,CANTATA:2021ktz}.

One such alternative relies on 3-form fields \cite{Koivisto:2009fb,Koivisto:2012xm}. These have been shown to yield accelerated solutions in cosmology, useful to explain both early (primordial inflation) and late (dark energy) time epochs of our universe \cite{Koivisto:2009ew,Morais:2016bev,Bouhmadi-Lopez:2016dzw}. Inflation under 3-form fields has been further studied in Ref.~\cite{Bouhmadi-Lopez:2016dzw,Germani:2009iq,Mulryne:2012ax,DeFelice:2012jt,Kumar:2014oka,Barros:2015evi,SravanKumar:2016biw}. More generally, they are also useful to describe the geometry of stars, black holes, wormholes or singularities \cite{Barros:2018lca,Barros:2020ghz,Bouhmadi-Lopez:2020wve,Barros:2021jbt,Bouhmadi-Lopez:2021zwt,Morais:2017vlf,Bouhmadi-Lopez:2018lly}, as well as Bianchi cosmologies \cite{Normann:2017aav} or extra-dimensional branes \cite{Barros:2023nzr}. The general formalism of 3-forms has been developed in previous studies such as those in Refs.~\cite{Koivisto:2009ew,Wongjun:2016tva,BeltranAlmeida:2018nin}.

While in four spacetime dimensions a standard canonical massive 3-form can be dualized into a scalar field model \cite{BeltranAlmeida:2018nin,Mulryne:2012ax,Germani:2009iq}, certain details warrant careful consideration. The dualization procedure can easily fail, depending on the physical nature one chooses to endow the theory with \cite{Duff:1980qv,Koivisto:2009fb}. For example, the introduction of a non-minimal coupling with other fields or the consideration of a more involved self-interacting potential for the 3-form are features known to break the mapping between the two frameworks \cite{SravanKumar:2016biw,Koivisto:2009sd}. Even in cases where the dualization is well defined, the scalar field representation often proves intractable, while the 3-form description remains effortless and intuitive \cite{Koivisto:2009fb}. Thus, the relevance of theories involving 3-forms can manifest itself in one of two scenarios. In either case, the outcome of the dualization process determines our trajectory: either we study what might be considered an intricate non-canonical scalar field theory, or the dualization process proves unsuccessful (breaks), leading us to investigate a novel cosmological theory centered on massive 3-forms. In either scenario, it is worth exploring the potential for emergent physics within these frameworks.

The conventional model of a massive canonical 3-form field, with a canonical kinetic term and for certain self-interacting potentials, has been shown to be incapable of generating scaling solutions \cite{Koivisto:2009fb,Koivisto:2009ew}. Nevertheless, in Ref.~\cite{Wongjun:2016tva} the author succeeded in deriving a Lagrangian density for a 3-form that mimics the cosmological evolution similar to that of a pressureless matter component with ${\rho\propto a^{-3}}$. This simply manifests as a specific kinetic 3-form term in the action. If one considers this Lagrangian term to be dominant during the early epochs, one can conceive of this theory as a 3-form dark energy model that effectively scales with pressureless matter at the outset. It only remains for a self-interaction term to dominate in later stages, successfully driving the current acceleration phase with an equation of state $w<-1/3$. In Ref.~\cite{Wongjun:2016tva}, the author suggested this possibility by means of an exponential self-interaction term. In this work, we implement this idea and explore its consequences.

We consider these results as the basic motivation for the present work and study such a dark energy model in detail. We derive an original Lagrangian in section \ref{sec:Lagrangian}. The analysis of the associated dynamical system in section \ref{sec:dynamical_system} allows us to study the cosmological dynamics in the presence of the 3-form field and a matter fluid. The implications for cosmology are highlighted in section \ref{sec:cosmology} before concluding in section \ref{sec:conclusion}.

\section{The Lagrangian}
\label{sec:Lagrangian}

In this section, we derive a comprehensive Lagrangian density that endows the 3-form with a specific dynamical behavior similar to that of a fluid with a constant equation of state $w$. While this result was mentioned in Ref.~\cite{Wongjun:2016tva}, we have uncovered a more general solution that not only complements the focus of that work, but also serves as a guide for the construction of a novel Lagrangian. This allows us to identify the particular terms in the action that initially induce the fluid behavior of either radiation or matter. Using this as a starting point, we then introduce a self-interaction term that will govern the cosmological dynamics at later times, responsible for the accelerated expansion.

Let us start by considering a general action for a single 3-form field \textbf{A}, minimally coupled to Einstein's gravity,
\be
\mathcal{S}=\int\dd^4 x \sqrt{-g}\left[ \frac{R}{2\kappa^2}+P(K,Y)\right]\,.
\ee
In the above equation, ${g}$ stands for the determinant of the metric ${g_{\mu\nu}}$, ${R}$ is the curvature scalar, ${\kappa^2=8\pi G}$ and $P$ is a general function of the Lorentz invariants:
\begin{eqnarray}
    K &=& -\frac{1}{48}F^2:=-\frac{1}{48}F_{\mu\nu\alpha\beta}{F}^{\mu\nu\alpha\beta}\,, \\
Y&=&\frac{1}{12}{A}^2 := \frac{1}{12}A_{\mu\nu\alpha}A^{\mu\nu\alpha}\,,
\end{eqnarray}
with ${\textbf{F}=\rm{d}\textbf{A}}$
representing the anti-symmetric strength tensor of the 3-form, with components ${F_{\mu\nu\alpha\beta}=4\partial_{\,[\mu}A_{\nu\alpha\beta]}}$. The equations of motion for the 3-form derived from such action are
\be\label{eq:eom}
\nabla_\mu \left(\frac{\partial P}{\partial K} F^\mu{}_{\alpha\beta\gamma}\right) = -\frac{\partial P}{\partial Y} A_{\alpha\beta\gamma}\,,
\ee
which can be equivalently derived from the conservation of the energy-momentum tensor, which has components
\be\label{em_tensor}
T_{\mu\nu}=\frac{1}{6}\frac{\partial P}{\partial K}{F}_{\mu\alpha\beta\gamma}{F}_{\nu}{}^{\alpha\beta\gamma}-\frac{1}{2}\frac{\partial P}{\partial Y}{A}_{\mu\alpha\beta}{A}_{\nu}{}^{\alpha\beta}+Pg_{\mu\nu}\,.
\ee
In the expression above we identify the energy density and pressure of the 3-form as,
\begin{eqnarray}
\rho&=&2K\frac{\partial P}{\partial K}-P\,,\label{rho}\\
p &=& -\rho-2Y\frac{\partial P}{\partial Y}\,,\label{pressure}
\end{eqnarray}
yielding its equation of state parameter,
\be
w=\frac{p}{\rho}=-1-\frac{2Y}{\rho}\frac{\partial P}{\partial Y}\,.
\ee
Rearranging Eq.~\eqref{eq:eom}, making use of Eqs.~\eqref{rho} and \eqref{pressure}, and introducing $\gamma\equiv1+w$ as the adiabatic index of the 3-form fluid, we have the following partial differential equation,
\be
2Y\frac{\partial P}{\partial Y}+\gamma\,\left(2K\frac{\partial P}{\partial K}-P\right)=0\,,
\ee
which has the following analytical solution for a constant index $\gamma$,
\be
\label{eq:general_function}
P=Y^{\gamma/2}f\left(\frac{K}{Y^{\gamma}}\right)\,,
\ee
where $f$ is a general function of the argument inside brackets. This solution generalizes the result of Ref.~\cite{Wongjun:2016tva}, which found the subset of solutions consisting of the identity function $f(x)=x$.

The behavior of a cosmological constant ($\Lambda$), pressureless component like cold dark matter (CDM), and radiation can be realized with ${\gamma=0}$, ${\gamma=1}$, and ${\gamma=4/3}$, respectively. It is well known that a massless 3-form, ${P\equiv P(K)}$, mimics a cosmological constant \cite{Hawking:1984hk}, which is consistent with Eq.~\eqref{eq:general_function} for ${\gamma=0}$. S. Hawking and N. Turok adopted this concept to tackle the cosmological constant problem \cite{Turok:1998he}. For the case ${\gamma=1}$, the 3-form effectively describes a non-relativistic fluid, since it can be shown that ${c_s^2=0}$. We report here new exact Lagrangians that mimic the behavior of $\Lambda$+CDM at the background level.

In order to concretely compute cosmological observables, it is necessary to choose a specific 3-form model. Throughout this work we consider the simplest case of Eq.~\eqref{eq:general_function}, which depends linearly on $K$ \cite{Wongjun:2016tva}. To identify the 3-form with a dynamical dark energy field, we also need to add a component that dominates the background density at late times. This can be done by including a self-interaction term, $V=V(Y)$, similar to the potential energy in standard scalar field cosmology. We end up with the following 3-form Lagrangian:
\be\label{eq:lag_gen}
\mathcal{L}_A=\frac{2K}{(2Y)^{\gamma/2}}-V\,,
\ee
where the factor $2$ simplifies the equations from now on. The form of the Lagrangian \eqref{eq:lag_gen} derives from a phenomenological approach, but the action can be framed in an effective field theory. The latter would be valid in some low-energy range of a more fundamental high-energy model, where non-canonical 3-form fluxes can emerge after dimensionality reduction \cite{Cremmer:1978km,Avetisyan:2022zza,Evnin:2023ypu,Vanichchapongjaroen:2020wza,Nitta:2018yzb}.

In the following two sections, we will particularize this Lagrangian to perform the analysis of the dynamical system (derived from the main cosmological equations) and the corresponding 3-form cosmology. Our main goal is to consider a 3-form dark energy Lagrangian with a dust-like energy density, ${\gamma=1}$, which eventually freezes. The acceleration of the universe at late times  is achieved by choosing an exponential form for the interaction term. Although this choice is again phenomenological, we consider self-interactions of more fundamental fields. In superstring cosmology, it has also been shown that an exponential potential can stabilize the dilaton \cite{Barreiro:1998aj}.

For now, we derive the cosmological background equations for a general and constant index $\gamma$, and an unspecified potential $V$. Let us consider a flat Friedmann–Lema\^{i}tre–Robertson–Walker (FLRW) line element, which describes an homogeneous and isotropic expanding universe:
\be
\dd s^2 = -\dd t^2 + a(t)^2\delta_{ij}\dd x^i \dd x^j\,,
\ee
where $a(t)$ is the scale factor, which is a function of cosmic time $t$. The most general parametrization for the 3-form, compatible with the FLRW symmetries, is given by \cite{Germani:2009iq},
\be
\label{3_form_flrw}
A_{ijk}=a^3\epsilon_{ijk}\chi(t)\,,
\ee
where $\chi$ is a scalar function that fully determines the 3-form components. The contractions of the kinetic and field terms are now
\be
\label{eq:contractions}
A^2 = 6\chi^2\quad\text{and}\quad F^2 = -24\left(\dot{\chi}+3H\chi\right)^2\,,
\ee
respectively, where a dot denotes a derivative with respect to cosmic time and we have defined the Hubble rate ${H=\dot{a}/a}$. Considering the general Lagrangian Eq.~\eqref{eq:lag_gen}, the equation of motion Eq.~\eqref{eq:eom} for the 3-form 
in terms of $\chi$, in an FLRW background, is,
\be\label{eom_chi}
\ddot{\chi}+\dot{\chi}\left(3H-\gamma\,\frac{\dot{\chi}}{2\chi}\right)+3\chi\left(\dot{H}+\frac{3}{2}\gamma\,H^2\right)+\frac{\chi^{\gamma}\,V_{,\chi}}{2}=0\,,
\ee
where ${V_{,\chi}=\partial V/\partial\chi}$. The energy density and pressure of the 3-form can be found through the time and spatial components of Eq.~\eqref{em_tensor}, 
\ba
\label{eq:rho_p}
\rho_\chi &=& \frac{\left(\dot{\chi}+3H\chi\right)^2}{\chi^\gamma}+V\,,\\
p_\chi &=& \left(\gamma-1\right)\frac{\left(\dot{\chi}+3H\chi\right)^2}{\chi^\gamma}-V +\chi V_{,\chi}\,,
\ea
respectively. Consistently, for ${\gamma=1}$ and in the absence of potential, ${V=0}$, the pressure vanishes. 

In this work, we focus on late-time cosmology. Therefore, we consider a universe filled with the 3-form fluid and a matter component consisting of baryons and cold dark matter. The field equations then give the Friedmann constraint, which describes the evolution of the Hubble rate, proportional to the energy densities,
\be
\label{eq:friedmann}
\frac{3}{\kappa^2}H^2 = \rho_\chi+\rho_m\,.
\ee
Taking the time derivative of the above equation and making use of Eq.~\eqref{eom_chi}, we get the Raychaudhuri equation,
\be
\label{eq:H_dot}
-\frac{2}{\kappa^2}\dot{H}=\rho_m+\gamma\,\frac{\left(\dot{\chi}+3H\chi\right)^2}{\chi^\gamma}+\chi V_{,\chi}\,,
\ee
where matter follows the standard continuity equation ${\dot{\rho}_m=-3H\rho_m}$.

At this point, it is possible to solve the background cosmology numerically. Nevertheless, in order to shed light on the overall behavior of the solutions, we employ dynamical system analysis techniques. This approach enables us to identify critical values for the variables and parameters that exert a significant influence on the phase space of the solutions, thereby exploring all potential solutions for the system.

\section{Dynamical system analysis}
\label{sec:dynamical_system}

\begin{table*}[]
\renewcommand{\arraystretch}{2} 
\setlength{\tabcolsep}{5pt} 
\centering
\begin{tabularx}{\linewidth}{|>{\centering\arraybackslash}X|>{\centering\arraybackslash}p{2.1cm}|>{\centering\arraybackslash}p{1.9cm}|>{\centering\arraybackslash}p{2.1cm}|>{\centering\arraybackslash}X|>{\centering\arraybackslash}X|>{\centering\arraybackslash}X|>{\centering\arraybackslash}X|>{\centering\arraybackslash}X|}
\hline
\textbf{Point} & $x$ & $y$ & $v^2$ & Existence & Stability & $\Omega_\chi$ & $w_\chi$ & $w_{\rm eff}$ \\ \hline
 $\bold{(A)}$ & $x$ & $1$ & $0$ & Always & Saddle & $x$ & 0 & 0 \\
 $\bold{(B)}$ & 0 & 0 & 0 & Always & Saddle & 0 & $-$ & 0 \\
 $\bold{(C)}$ & 0 & 0 & 1 & Always & \textbf{Stable} & 1 & $-1$ & $-1$ \\
 $\bold{(D)}$ & $\frac{1}{2}\left(1-\sqrt{1-\lambda}\right)$ & $1$ & $\frac{1}{2}\left(1+\sqrt{1-\lambda}\right)$ & $0<\lambda<1$ & Saddle & $1$ & $-1$ & $-1$\\
 $\bold{(E)}$ & $\frac{1}{2}\left(1+\sqrt{1-\lambda}\right)$ & $1$ & $\frac{1}{2}\left(1-\sqrt{1-\lambda}\right)$ & $0<\lambda<1$ & \textbf{Stable} & $1$ & $-1$ & $-1$\\ \hline
\end{tabularx}
\caption{Critical points and their condition of existence.}
\label{tab:critical}
\end{table*}

The following sections focus on the adiabatic index ${\gamma=1}$, which corresponds to the pressureless 3-form case of interest. As mentioned above, we also consider an exponential potential, which is the simplest proposal in Ref.~\cite{Wongjun:2016tva},
\be
\label{eq:potential}
V=V_0\exp({-3A^2/\lambda})\,,
\ee
where $V_0$ is the mass scale, and $\lambda$ is a non-vanishing constant that conveniently parametrizes the slope of the potential. We write derivatives with respect to the number of $e$-folds, $N\equiv\ln a$, which simplifies the reading of the dynamical system. They are denoted by a prime, i.e, ${x'=\dd x/\dd\ln a}$. Additionally, without loss of generality, we set ${\kappa=1}$.

The equation of motion \eqref{eom_chi} for the scalar function, that fully determines the 3-form, can be rewritten as
\be\label{eom2_chi}
\chi''+\chi'\left(3+\frac{H'}{H}-\frac{\chi'}{2\chi}\right)+3\chi\left(\frac32+\frac{H'}{H}\right)-\frac{18\chi^2V}{\lambda H^2}=0\,.
\ee
Let us introduce the following set of variables:
\begin{align}
\label{eq:variables}
\begin{aligned}
x   &= 3\chi\,,           & y   &= \frac{\chi' + 3\chi}{3\chi}\,, \\
v^2 &= \frac{V}{3H^2}\,,  & z^2 &= \frac{\rho_m}{3H^2}\,.
\end{aligned}
\end{align}
From Eq.~\eqref{eq:rho_p}, the physical variables $xy^2$ and $v^2$ represent the fractional energy density of the kinetic and potential terms of the 3-form, respectively, while ${z^2=\Omega_m}$ is the fractional energy density of matter. The Friedmann equation \eqref{eq:friedmann} gives a constraint,
\be
\label{eq:constraint}
1=xy^2+v^2+z^2\,,
\ee
which can be used to eliminate one variable from the system. From the Raychaudhuri equation \eqref{eq:H_dot}, we can also write
\be\label{eq:HlH}
\frac{H'}{H}=-\frac32\left(xy^2+z^2-\frac{4v^2x^2}{\lambda}\right)\,.
\ee
The phase space of the 3-form dynamics is characterized by the following set of first order and  autonomous differential equations:
\ba
x'&=&3x\left(y-1\right)\,,\\
y'&=&-y\left(\frac{H'}{H}+\frac{3}{2} y\right)+\frac{6v^2x}{\lambda}\,,\\
v'&=& -v\left[\frac{H'}{H}+\frac{12x^2}{\lambda}\left(y-1\right)\right]\,,\\
z'&=&-z\left(\frac32+\frac{H'}{H}\right)\,.
\ea

The relative energy density and equation of state of the 3-form are
\ba
\Omega_\chi&=& 1-z^2\,,\\
w_\chi &=& -\frac{v^2}{xy^2+v^2}\left(1+\frac{4x^2}{\lambda}\right)\,,
\ea
respectively. Imposing that the energy density of the 3-form does not overclose the universe and is always positive implies ${0\leqslant\Omega_\chi\leqslant1}$, which is equivalent to
\be
0\leqslant xy^2+v^2\leqslant1\,.
\ee
The effective equation of state parameter,
\be
\label{eq:w_eff}
w_{\rm eff}=w_\chi\Omega_\chi=-v^2\left(1+\frac{4x^2}{\lambda}\right)\,,
\ee
can ascertain whether the universe undergoes an accelerated (${w_{\rm eff}<-1/3}$) or decelerated (${w_{\rm eff}>-1/3}$) expansion. It also allows us to rewrite Eq.~\eqref{eq:HlH} as
\be
\frac{H'}{H}=-\frac32\left( 1+w_{\rm eff} \right)\,.
\ee
Finally, by choosing to eliminate the variable $z$ through Eq.~\eqref{eq:constraint}, the dynamical system reads
\begin{subequations}
\label{eq:autonomous}
\begin{align}
x' &= 3x\left(y-1\right)\,,\\
\label{eq:x_prime}
y' &=-\frac{3}{2}y\left(v^2+y-1\right)-\frac{6v^2x}{\lambda}\left(xy-1\right)\,,\\
\label{eq:y_prime}
v' &= -\frac{3v}{2\lambda}\left[4x^2\left(v^2+y-1\right)+\left(v^2-1\right)\lambda\right]\,.
\end{align}
\end{subequations}

As the kinetic and field term contractions, described in Eq.~\eqref{eq:contractions}, are invariant under the transformation ${\chi\rightarrow-\chi}$, we limit our analysis to the case ${x>0}$. Furthermore, we restrict it to ${v>0}$ as the dynamical system possesses the symmetry ${v\rightarrow-v}$.

The fixed points play a crucial role in characterizing the steady-state configurations of the system. These points correspond to critical configurations of the background cosmology, providing significant insights into its behavior. These are reported in Table~\ref{tab:critical} and are identified by solving the polynomial equations resulting from setting the left-hand side of the autonomous equations \eqref{eq:autonomous} to zero. For each equilibrium point, we provide the corresponding cosmological parameters, namely $\Omega_\chi$, $w_\chi$ and $w_{\rm eff}$, to facilitate a thorough examination of the 3-form cosmology in the subsequent section.

To investigate the stability of the system, we study the trajectories in the vicinity of the fixed points. Small perturbations are considered around them in order to define the stability matrix, from which the eigenvalues, ${(e_1,e_2,e_3)}$ listed in Appendix \ref{appendix_table_eigen}, are derived. These values allow for the determination of the stability of the critical points, which can then be classified into three distinct categories: stable, unstable, or saddle points. We find that the behavior of the dynamical system is contingent upon the values of the 3-form potential slope.

\begin{figure}[t]
\rotatebox{-6.5}
{\includegraphics[height=1\linewidth]{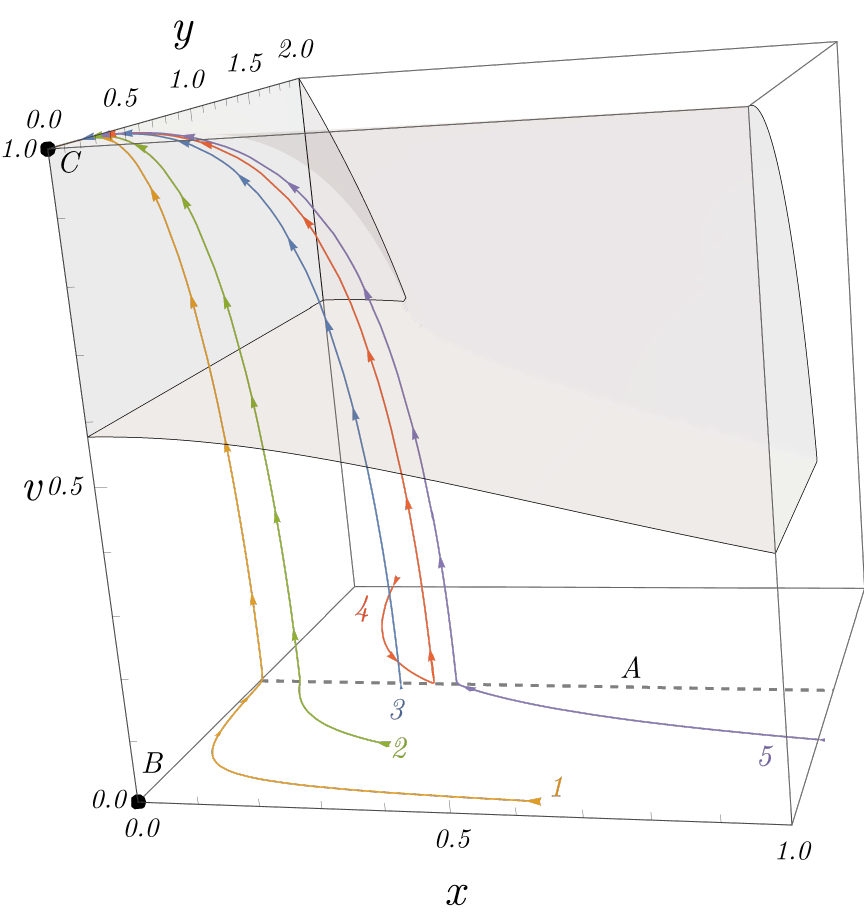}}
\captionsetup{justification=raggedright,singlelinecheck=false}
  \caption{Phase
portrait for $\lambda=3.6$ and different initial conditions. The shaded volume corresponds to the region where the expansion accelerates. The dashed line is the critical point $A$, i.e., the line $(y,v)=(1,0)$.}
  \label{fig:traj_small}
\end{figure}

\begin{figure}[t]
\rotatebox{-6.5}
{\includegraphics[height=1\linewidth]{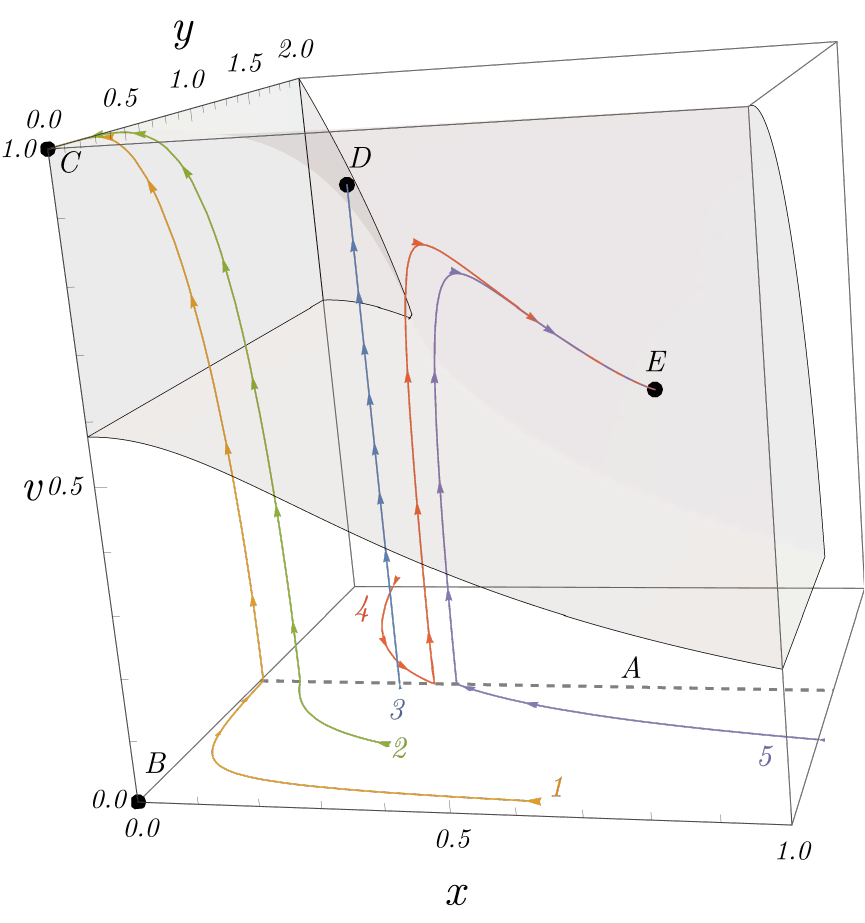}}
\captionsetup{justification=raggedright,singlelinecheck=false}
  \caption{Phase
portrait for $\lambda=0.8$ and different initial conditions. The shaded volume corresponds to the region where the expansion accelerates. The dashed line is the critical point $A$, i.e., the line $(y,v)=(1,0)$.}
  \label{fig:traj_large}
\end{figure}

\begin{itemize}
    \item \textbf{Point (A):}  This critical point always exists and corresponds to the 3-form being kinetically dominated. The 3-form behaves like a matter fluid ($w_\chi=0$), and there is no accelerated expansion. In terms of stability, it is a saddle point because eigenvalues have different signs along different directions.
    \item \textbf{Point (B):} This is the trivial solution of the system at the origin of the phase space. It always exists and corresponds to the scenario where the 3-form is absent in a universe completely dominated by matter. With eigenvalues of different signs, the critical point (B) is of saddle type. 
    \item \textbf{Point (C):} This equilibrium point of the system, which is always present, corresponds to a universe completely dominated and accelerated by the potential, where the 3-form acts as a cosmological constant (${w_\chi=-1}$). It has two negative eigenvalues ($e_1$ and $e_3$), and a vanishing one ($e_2$). Given the presence of a neutral eigenvalue, we have analyzed the center manifold for this point, as in Ref.~\cite{Boehmer:2011tp}, to confirm its stability. The dynamics of the system is determined by the evolution of the center mode, which indicate that the fixed point (C) is asymptotically stable (see Appendix \ref{appendix_center_manifold} for more details). It is therefore a non-hyperbolic attractor.
    \item \textbf{Point (D):} It is a solution that exists only for ${0<\lambda<1}$. It corresponds to an accelerated universe dominated by the 3-form. The potential is always larger than the kinetic term. The point is of saddle type, with two negative eigenvalues ($e_1$ and $e_2$) and one positive ($e_3$). 
    \item \textbf{Point (E):} This solution, which exists only for ${0<\lambda<1}$, corresponds to a universe in acceleration where the kinetic term of the 3-form is greater than its potential term. The critical point is stable, as evidenced by the negative value of the three corresponding eigenvalues, indicating that trajectories in phase space are attracted to this solution. Moreover, for ${0<\lambda<3/4}$, two eigenvalues acquire a complex nature with negative real parts, ${\Re(e_2)=\Re(e_3)=-3/2}$, while the imaginary parts induce an oscillatory behavior of the solutions. In this case the fixed point (E) is a stable spiral.
\end{itemize}

We now proceed to describe the possible trajectories in phase space over three different parametric ranges of $\lambda$, each of which exhibits different overall behaviors relevant to the present study.

\subsection{Case $\lambda\geqslant1$}

As shown in Fig.~\ref{fig:traj_small}, the critical point (C) is the only stable fixed point to which every trajectory converges, necessarily crossing into the acceleration region. For initial conditions where the 3-form potential is relatively small, i.e, ${v_i\approx0}$, the trajectories in the three-dimensional phase space are attracted and then repelled by the saddle points (A). These points are represented in Fig.~\ref{fig:traj_small} by a dashed gray line ${y=1}$ in the ${v=0}$ plane. For initial small values of $y$, they are first attracted and repelled by the saddle point (B), as illustrated by the orange orbit.

\subsection{Case $3/4\leqslant{\lambda<1}$}

We show in Fig~\ref{fig:traj_large} that the dynamical system has the same stable point (C) and the same saddle points (A) and (B), as in the case of large $\lambda$. However, since $\lambda<1$, there are two additional critical points, the stable point (E) and the saddle point (D). There are three possible dynamical behaviors that depend on the initial condition for the 3-form energy density, $\Omega_\chi^i$. Using \eqref{eq:constraint}, we can write
\be
\label{eq:negligeable_potential}
xy^2\approx\Omega_\chi^i\,,
\ee
as long as the 3-form potential is negligible.  The trajectories within the $v=0$ plane follow hyperbolic paths towards the critical (dashed) line (A) in Fig.\ref{fig:traj_large}, which intersects the $y$-plane at ${y=1}$. This means that, at $(v,y)\approx(0,1)$, we have
\be
\label{eq:x_A}
x \approx \Omega_\chi^i\,,
\ee
as seen in Table~\ref{tab:critical}. By equating this value with the $x$-value of the saddle point (D), we can obtain a critical fractional energy density for the 3-form as
\be
\label{eq:critical}
\Omega_\chi^c = \frac{1-\sqrt{1-\lambda}}{2}\,.
\ee
Thus, trajectories starting from negligible potential fractional energy, $v_i\approx 0$, evolve according to one of three possibilities. If ${\Omega_\chi^i<\Omega_\chi^c}$, then the trajectories follow paths close to the heteroclinic orbit $\mathrm{(A)}\rightarrow\mathrm{(D)}\rightarrow\mathrm{(C)}$ (orange orbit 1 and green orbit 2 in Fig.~\ref{fig:traj_large}). If ${\Omega_\chi^i>\Omega_\chi^c}$, they closely follow $\mathrm{(A)}\rightarrow\mathrm{(D)\rightarrow\mathrm{(E)}}$ (red orbit 4 and purple orbit 5 in Fig.~\ref{fig:traj_large}). Finally, for ${\Omega_\chi^i=\Omega_\chi^c}$, the solution reduces to the heteroclinic orbit $\mathrm{(A)}\rightarrow\mathrm{(D)}$ (blue orbit 3 in Fig.~\ref{fig:traj_large}). All these trajectories necessarily end in the acceleration region.

\subsection{Case ${0<\lambda<3/4}$}

\begin{figure}[h]
\rotatebox{-7}
{\includegraphics[height=1\linewidth]{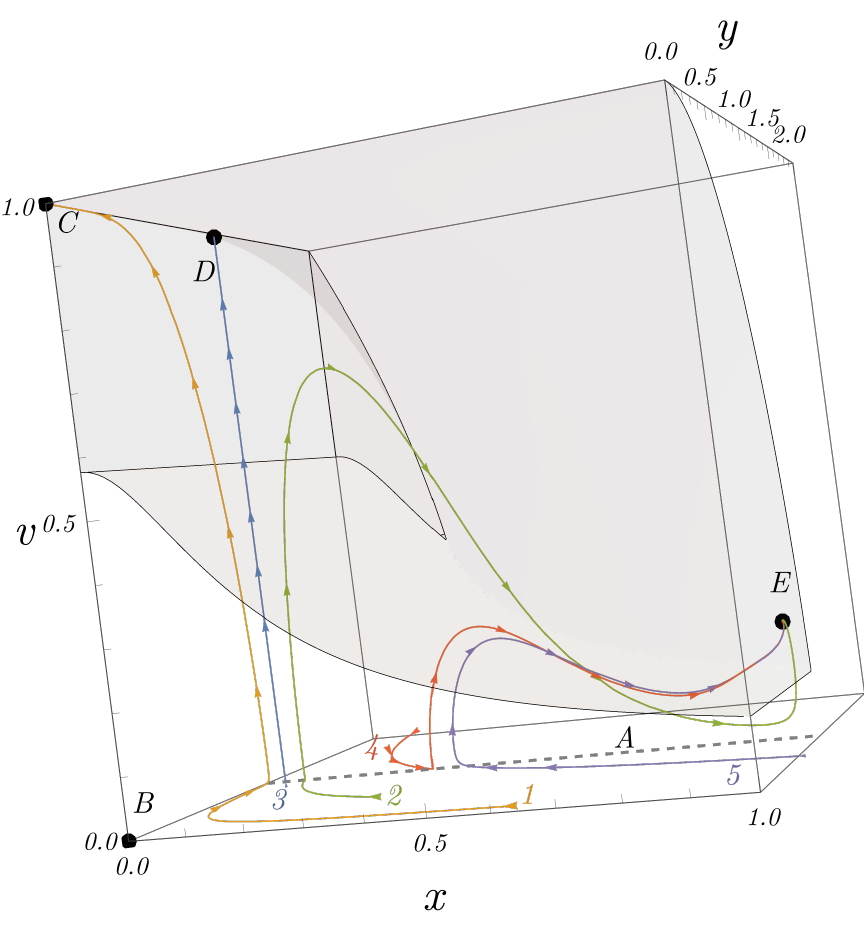}}
\captionsetup{justification=raggedright,singlelinecheck=false} 
  \caption{Phase
portrait for $\lambda=0.14$ and different initial conditions. The shaded volume corresponds to the region where the expansion accelerates. The dashed line is the critical point $A$, i.e., the line $(y,v)=(1,0)$.}
  \label{fig:traj_very_large}
\end{figure}

For these smaller values of the field potential parameter, the critical points and the overall behavior of the dynamical system are identical to the previous case where $3/4\leqslant{\lambda<1}$. However, the oscillations can cause the trajectories to cross the acceleration plane several times. This case is illustrated in Fig.~\ref{fig:traj_very_large}, where we observe that the oscillations in the green trajectory 2 cause it to cross the acceleration plane three times. Thus, the universe can have an acceleration phase, followed by a deceleration phase, followed by another acceleration phase. Typically, given the low frequency of the oscillations, we do not expect more than two acceleration periods. This opens up an interesting possibility akin to quintessential inflation models \cite{Peebles:1998qn}, which is beyond the scope of this study.

\begin{figure*}[t]
\includegraphics[height=0.65\linewidth]{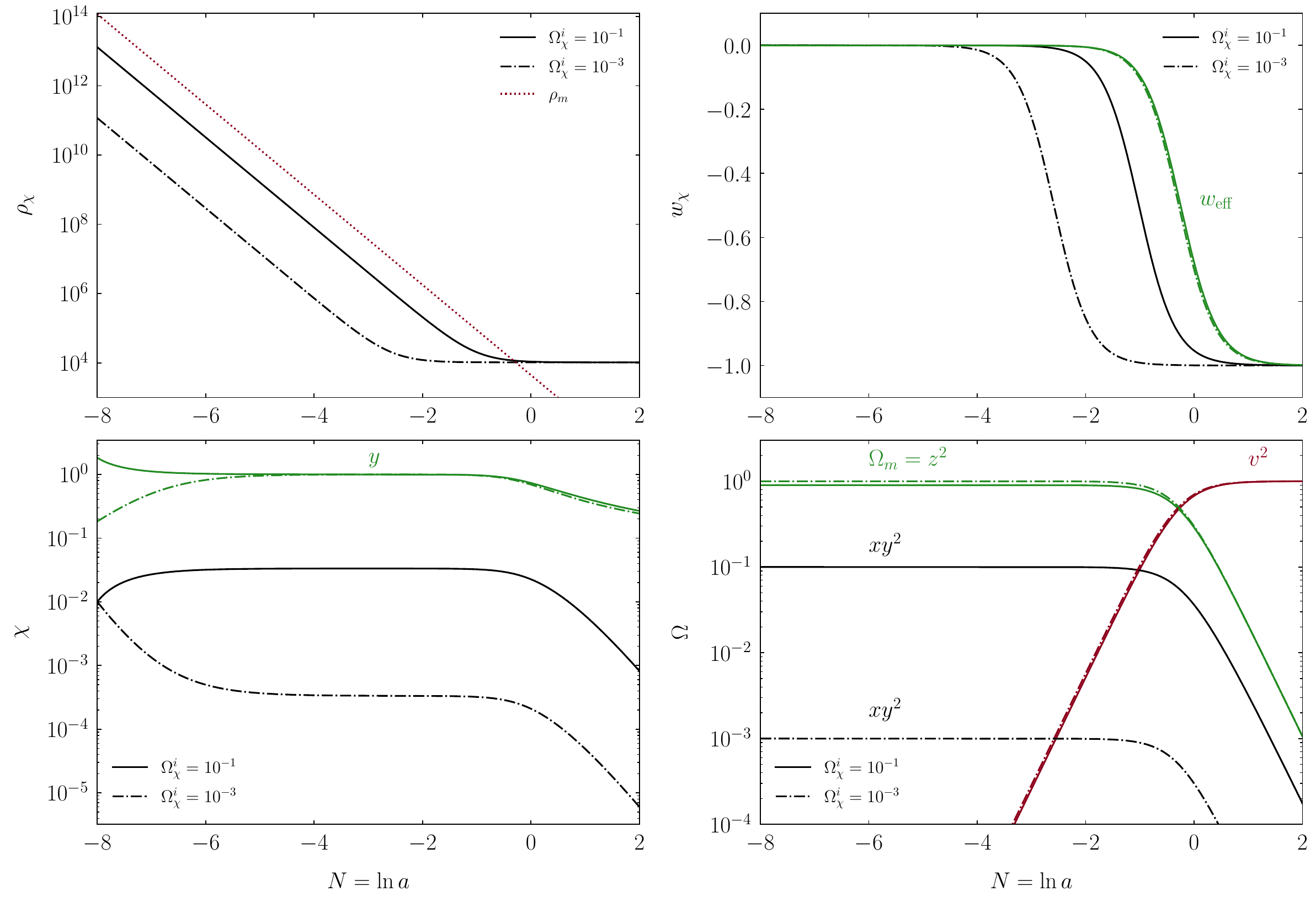}
\captionsetup{justification=raggedright,singlelinecheck=false}
    \caption{Cosmological evolution over the number of e-folds with $\lambda=3.6$, for two different initial 3-form abundances, $\Omega_\chi^i=10^{-1}$ and $\Omega_\chi^i=10^{-3}$. The initial condition of the field is arbitrarily set to $\chi_i=10^{-2}$. \underline{Upper left panel}: energy densities (3-form in black and matter in red). \underline{Upper right panel}: Equations of state (3-form in black and effective fluid in green). \underline{Lower left panel}: 3-form field in black and variable $y$ of Eq.~\eqref{eq:variables} in green. \underline{Lower right panel}: fractional energy densities (matter in green, field potential in red, and field kinetic energy in black).}
    \label{fig:small_lambda_value}
\end{figure*}

\section{3-form background cosmology}
\label{sec:cosmology}

The previous analysis of the dynamical system allows us to study the underlying cosmology of a universe composed of the 3-form and matter. We choose to start the evolution of the background at the e-folding time ${N_i=-8}$ in the numerical simulations. The initial conditions are denoted by subscript or superscript $i$.

For a given fractional energy density of the 3-form, $\Omega_\chi^i$, at the beginning of the evolution, the initial expansion rate is
\be
\label{eq:Hi}
H_i^2=\frac{1}{3}\frac{\rho_m^i}{1-\Omega_\chi^i}\,,
\ee
with
\be
\rho_{m}^i=3H_0^2\Omega_m^0\,e^{-3N_i}\,,
\ee
where $\Omega_m^0$ is the current abundance of matter, and $H_0$ is today's expansion rate. Using Eq.~\eqref{eq:potential}, we derive the very small initial condition for the fractional energy density of the 3-form potential, ${v_i^2=V_i/3H_i^2}$, as
\be
v_i^2 = \left(1-\Omega_\chi^i\right)\frac{1-\Omega_m^0}{\Omega_m^0}\,\exp{\left(-\frac{18}{\lambda}\chi_i^2+3N_i\right)}\,,
\ee
by assuming that the mass scale of the potential is of the order of a cosmological constant, ${V_0=3H_0^2(1-\Omega_m^0)}$. For a given initial fractional energy density of the 3-form $\Omega_\chi^i$ and a given initial field value $\chi_i$, we can find its corresponding initial velocity $\chi_i^\prime$, using Eq.~\eqref{eq:rho_p}:
\be
\frac{\left(\chi^\prime_i+3\chi_i\right)^2}{3\chi_i}=\Omega_\chi^i-v_i^2\,,
\ee
where the l.h.s of the above equation is the initial fractional energy density of the 3-form kinetic term.

The 3-form  dilutes like pressureless matter as the universe expands, as seen in the upper left panel of Fig.~\ref{fig:small_lambda_value}. At this point, we have that ${w_{\rm eff}=w_\chi=0}$ as seen in the upper right panel of Fig.~\ref{fig:small_lambda_value}. When the saddle point (A) of the dynamical system is reached, the $\chi$ field stabilizes on a plateau whose value can be obtained using Eq.~\eqref{eq:x_A}:
\be
\chi=\frac{x}{3}=\frac{\Omega_\chi^i}{3}\,.
\ee
This plateau depends only on the initial abundance of the 3-form, $\Omega_\chi^i$. We could therefore start the numerical evolution of the background close to this attractor by setting ${\chi_i=\Omega_\chi^i/3}$ and $\chi_i'=0$.

\begin{figure*}[t]
    \centering
\includegraphics[height=0.65\linewidth]{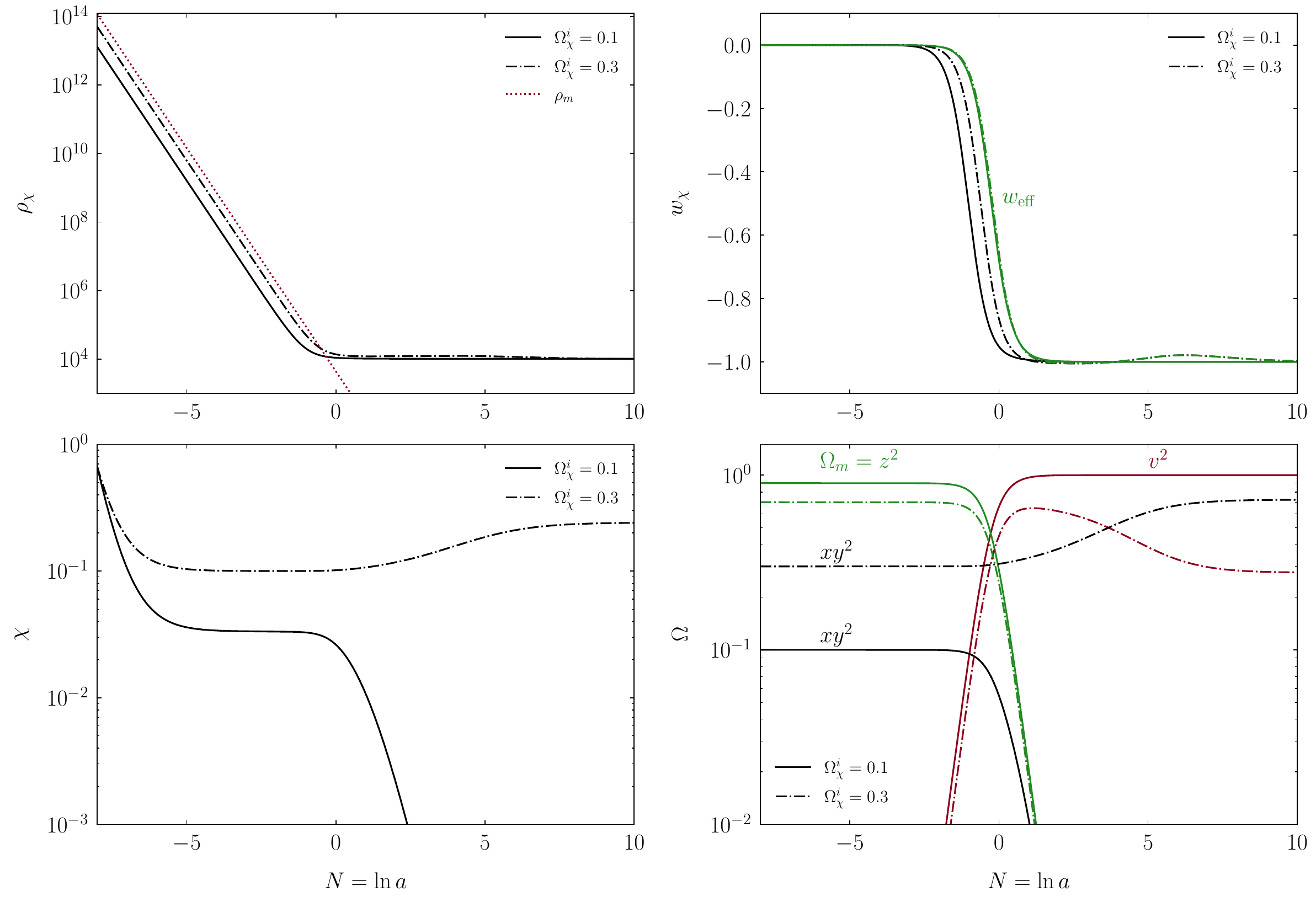}
\captionsetup{justification=raggedright,singlelinecheck=false}
    \caption{Cosmological evolution over the number of e-folds with $\lambda=0.8$, for two different initial ratios of the energy densities, $\Omega_\chi^i=0.1$ and $\Omega_\chi^i=0.3$. The initial condition of the field is arbitrarily set to $\chi_i=2/3$. \underline{Upper left panel}: energy densities (3-form in black and matter in red). \underline{Upper right panel}: Equations of state (3-form in black and effective fluid in green). \underline{Lower left panel}: 3-form field. \underline{Lower right panel}: fractional energy densities (matter in green, field potential in red, and field kinetic energy in black).}
\label{fig:large_lambda_value}
\end{figure*}

According to the analysis of the dynamical system in the previous section, for ${\lambda\geqslant1}$, every trajectory is eventually attracted to the stable point (C), where the field vanishes, as depicted in the lower left panel of Fig.~\ref{fig:small_lambda_value}. Consequently, we see in the lower right panel that the 3-form kinetic term, $xy^2$, becomes negligible and its potential dominates the universe lately with $v^2\approx1$. This corresponds to a scenario where the 3-form energy density resembles that of a cosmological constant with ${w_{\rm eff}=w_\chi=-1}$.

For ${\lambda<1}$ and ${\Omega_\chi^i>\Omega_\chi^c}$ in Eq.\eqref{eq:critical}, the future attractor is the stable point (E). In such a case, the field will eventually freeze at a value that depends only on $\lambda$,
\be
\chi= \frac{1}{6}\left(1+\sqrt{1-\lambda}\right)\,,
\ee
as illustrated in Fig.~\ref{fig:large_lambda_value}. The universe is finally dominated by the 3-form, which accelerates its expansion with ${w_{\rm eff}=w_\chi=-1}$. The kinetic energy of the field,
\be
xy^2=\frac{1}{2}\left(1+\sqrt{1-\lambda}\right)\,,
\ee
is larger than the potential energy,
\be
\label{eq:final_v}
v^2= \frac{1}{2}\left(1-\sqrt{1-\lambda}\right)\,.
\ee

We thus confirm that the solutions always evolve towards an accelerating phase of the expansion, regardless of the value of the potential slope.

\begin{figure*}[t]
    \centering
\includegraphics[height=0.65\linewidth]{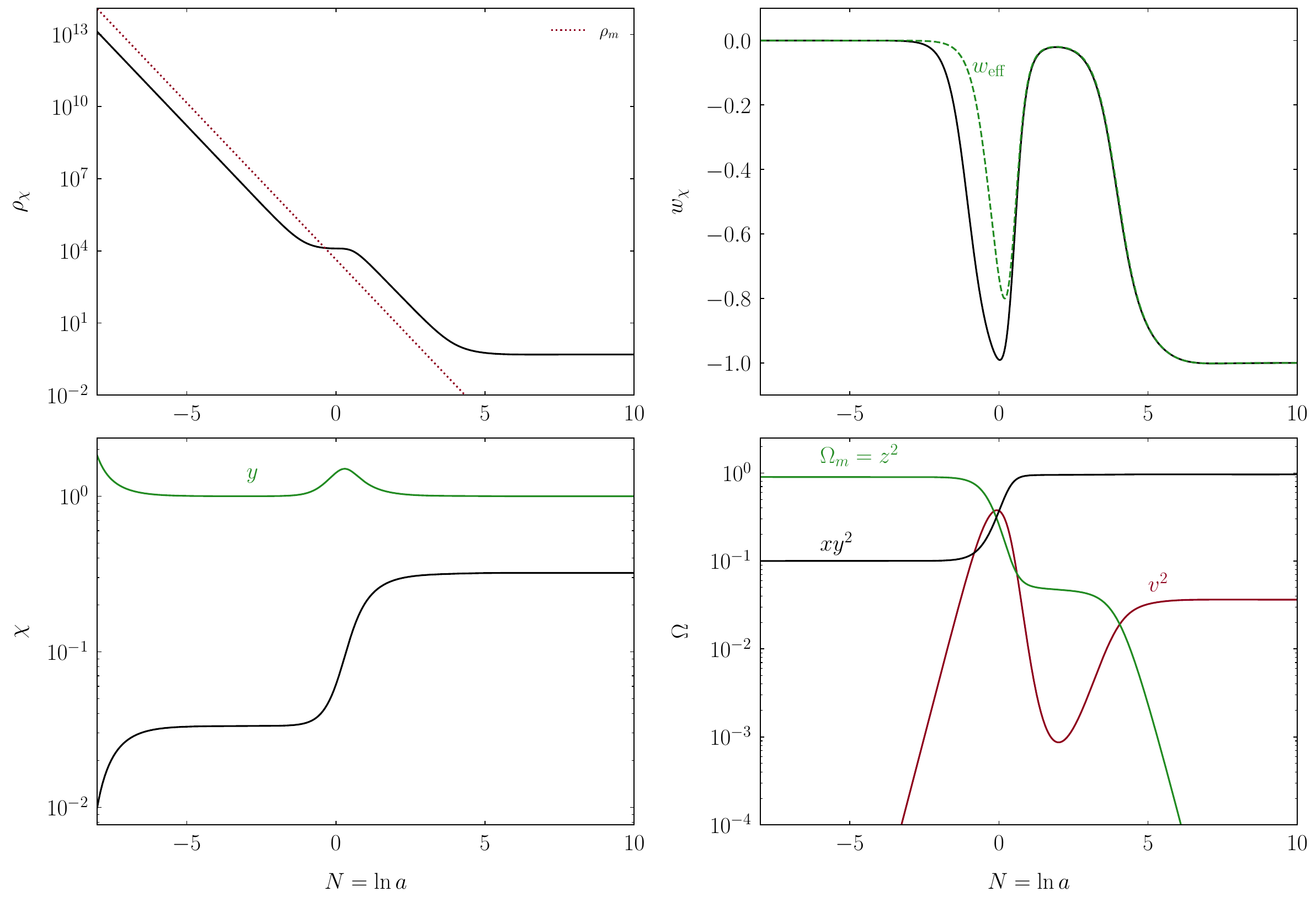}
\captionsetup{justification=raggedright,singlelinecheck=false}
    \caption{Cosmological evolution over the number of e-folds with $\lambda=0.14$ and $\Omega_\chi^i=0.1$. The initial condition of the field is arbitrarily set to $\chi_i=10^{-2}$. \underline{Upper left panel}: energy densities (3-form in black and matter in red). \underline{Upper right panel}: Equations of state (3-form in black and effective fluid in green). \underline{Lower left panel}: 3-form field in black and variable $y$ of Eq.~\eqref{eq:variables} in green. \underline{Lower right panel}: fractional energy densities (matter in green, field potential in red, and field kinetic energy in black).}
\label{fig:very_large_lambda_value}
\end{figure*}

Finally, there is a peculiar case with further implications for cosmology. When ${\lambda<3/4}$, the fractional energy of the potential can oscillate before settling down to its final value in Eq.~\eqref{eq:final_v}. Consequently, the trajectory in the phase space may leave the acceleration region before re-entering it. In this scenario, the universe temporarily ceases to experience acceleration as the 3-form energy density is once again diluted  by the expansion. Eventually, as the potential flattens, the 3-form energy density reaches a steady state, and the universe resumes its period of accelerated expansion. This scenario is depicted in Fig.~\ref{fig:very_large_lambda_value}. It is noteworthy that this mechanism can lead to the generation of multiple phases of acceleration.

\FloatBarrier

\section{Conclusion}
\label{sec:conclusion}

In this work, we explored the suggestions of Ref~\cite{Wongjun:2016tva}, introducing a generalized solution for describing a 3-form field \textbf{A} that behaves as a cosmological fluid with a constant equation of state parameter. We also developed a viable cosmological model by formulating a Lagrangian that endows the 3-form with the dynamical properties of both dust at the beginning, and dark energy accelerating the universe in late times. The acceleration is achieved by including a self-interacting exponential term of the form ${V\propto \exp(-3A^2/\lambda)}$, which stabilizes the solutions in an inflationary state. Resorting to dynamical systems techniques, we thoroughly analyzed the solutions considering a flat FLRW background universe filled with matter and the 3-form fluid. The latter, due to homogeneity and isotropy, can be fully parametrized in terms of a scalar, $\chi$, function of cosmic time only.

By analyzing the dynamical system composed of the equation of motion and the Friedmann constraint, we show that the overall behavior of the solutions depends only on the slope of the potential, parametrized by the constant $\lambda$, and the initial abundance of the 3-form fluid, $\Omega_\chi^i$. The system has one non-hyperbolic stable point, whose stability was inferred from the analysis of the center manifold. 

Solutions describing a late-time De Sitter expansion naturally arise. The accelerating regime is achieved regardless of the initial value and kinetic energy of the field. While the 3-form dilutes as dust throughout the first part of the evolution, its energy density eventually freezes, resembling a cosmological constant. As the effective equation of state parameter falls below $-1/3$, the expansion accelerates, and the 3-form assumes its role as dark energy.

For ${\lambda\geqslant1}$, the density of the 3-form is dominated by the potential energy, while for ${\lambda<1}$ it is the kinetic energy of the 3-form that may end up dominating. We have also shown that a stiff potential with ${\lambda<3/4}$ induce oscillations in the system. These solutions present an intriguing scenario, as they can result in multiple phases of inflation, interspersed with periods in which the expansion decelerates. These solutions may be effectively tailored to describe both primordial inflation and dark energy, with a subdominant period in between, resorting to a single 3-form. Such behavior was achieved with a scalar field in Ref.~\cite{Rosa:2019jci}. Thus, this theory can serve as a proxy for quintessential inflationary models with rich physical implications.

We can conclude that the shape of the Lagrangian allows a reasonable reconstruction of the cosmic history of the universe's background. Although we have focused solely on a cosmological subset of (non-canonical) 3-form-driven dark energy, the overall theory suggests much broader applications, such as in the early universe (primordial inflation) or oscillating universe. The theory may have the potential to drive both primordial inflation and late-time dark energy phenomena effectively with a single 3-form field. In this context, it may be interesting to further the case for an early radiation-like fluid with the adiabatic index ${\gamma=4/3}$.

For future work, it would be worthwhile to test the model against observational data. In particular, data sets of distance measurements, e.g. of supernovae and baryonic acoustic oscillations, could have the potential to constrain the model, which is subject to the usual fine-tuning of dark energy models, without having to go to the perturbation level. However, the latter will be necessary to test the model with cosmic microwave background and large-scale structure observables, probing high and low redshifts, respectively.

\acknowledgments
This work is supported by Funda\c{c}\~{a}o para a Ci\^{e}ncia e a Tecnologia (FCT) through the research grants DOI: 10.54499/UIDB/04434/2020 and DOI: 10.54499/UIDP/04434/2020, and through the BEYLA project, 10.54499/PTDC/FIS-AST/0054/2021. VdF acknowledges support from FCT through grant 2022.14431.BD.

\begin{table*}[t]
\renewcommand{\arraystretch}{2} 
\setlength{\tabcolsep}{5pt} 
\centering
\begin{tabularx}{\linewidth}{|>{\centering\arraybackslash}X|>{\centering\arraybackslash}X|>{\centering\arraybackslash}X|>{\centering\arraybackslash}X|}
\hline
 \textbf{Point}& $e_1$ & $e_2$ & $e_3$\\ \hline
 $\bold{A}$ & $0$ & $-3/2$ & $3/2$ \\
 $\bold{B}$ & $-3$ & $3/2$ & $3/2$ \\
 $\bold{C}$ & $-3$ & $0$ & $-3$ \\
 $\bold{D}$ & $-3$ & $-\frac{3}{2}\left(1+\sqrt{1+2\sqrt{1-\lambda}}\right)$ & $-\frac{3}{2}\left(1-\sqrt{1+2\sqrt{1-\lambda}}\right)$\\
 $\bold{E}$ & $-3$ & $-\frac{3}{2}\left(1+\sqrt{1-2\sqrt{1-\lambda}}\right)$ & $-\frac{3}{2}\left(1-\sqrt{1-2\sqrt{1-\lambda}}\right)$  \\ \hline
\end{tabularx}
\caption{Critical points and their eigenvalues.}
\label{tab:critical_appendix}
\end{table*}
\appendix
\section{Eigenvalues of the equilibrium points}\label{appendix_table_eigen}

In Table~\ref{tab:critical_appendix}, we list the eigenvalues ($e_1,e_2,e_3$) of the dynamical system for each critical point described in section~\ref{sec:dynamical_system}.


\section{Center manifold analysis}\label{appendix_center_manifold}

In this appendix we summarize the center manifold analysis which confirms the stability of the fixed point (C), with coordinates $(0,0,1)$ and eigenvalues $(-3,0,-3)$. The analysis follows the methodology described in Ref.~\cite{Boehmer:2011tp} .

The eigenvector matrix corresponding to point (C) is as follows
\[E=
\begin{pmatrix}
-\frac{\lambda}{2} & 1 & 0 \\
0 & 1 & 0 \\
0 & 0 & 1
\end{pmatrix}
.
\]
We introduce a new set of variables to move the critical point to the phase space origin. Let
\begin{align}
    X &= x\,, \nonumber\\
    Y &= y\,, \nonumber\\
    V &= v-1\,.\nonumber
\end{align}

The dynamical system can be decomposed into central and stable components, by a Jordan transformation from the state vector $(X,Y,V)$ to the new coordinates $(s,u,t)$ such that
\[
\begin{pmatrix}
s \\
u \\
t
\end{pmatrix}
=
\begin{pmatrix}
-\frac{2}{\lambda} & 0 & 0 \\
\frac{2}{\lambda} & 1 & 0 \\
0 & 0 & 1
\end{pmatrix}
\begin{pmatrix}
X \\
Y \\
V
\end{pmatrix}
,
\]
where the transformation matrix is the inverse of the transposed eigenvector matrix, $E$. In these coordinates, the dynamical system has the following form
\[
\begin{pmatrix}
s' \\
u' \\
t'
\end{pmatrix}
=
\begin{pmatrix}
-3 & 0 & 0 \\
0 & 0 & 0 \\
0 & 0 & -3
\end{pmatrix}
\begin{pmatrix}
s \\
u \\
t
\end{pmatrix}
+ \left(\vphantom{\begin{pmatrix} s' \\ u' \\ t' \end{pmatrix}} \text{non-linear terms} \right).
\]

The dynamics of the central component is
\begin{equation}
\label{eq:u'}
    u'= 2x'/\lambda+y'\,,
\end{equation}
where $x'$ and $y'$ respect the equations of the original referential, Eq.~\eqref{eq:x_prime} and Eq.~\eqref{eq:y_prime}, respectively. As for the stable modes, $s$ and $t$, they follow passively as functions of the variable $u$. The  corresponding functions, $h_s$ and $h_t$, can be approximated by polynomials for a desired degree of accuracy,
\begin{align}
    h_s &= s_2 u^2 + s_3 u^3 + \mathcal{O}(u^4)\,, \\
    h_t &= t_2 u^2 + t_3 u^3 + \mathcal{O}(u^4)\,.
\end{align}
Using the original equations \eqref{eq:autonomous} of the dynamical system, and substituting the polynomial functions into Eq.~\eqref{eq:u'}, we obtain,
\begin{equation}
\label{eq:u}
    u'=-\frac{3}{2}u^2+\mathcal{O}(u^3)\,,
\end{equation}
which indicates that the non-hyperbolic point (C) is stable. Note that the leading term in Eq.~\eqref{eq:u} does not depend on the coefficients of the polynomials $h_s$ and $h_t$.

\newpage
\bibliography{bib}

\end{document}